\documentclass[twocolumn,showpacs,showkeys,preprintnumbers,amsmath,amssymb,prb]{revtex4}

\usepackage{graphicx}

\begin{document}

\preprint{Wurmehl et al, Co$_2$FeSi, PRB}

\title{Geometric, electronic, and magnetic structure of Co$_2$FeSi:
       Curie temperature and magnetic moment measurements and calculations.}

\author{Sabine Wurmehl, Gerhard H. Fecher, Hem Chandra Kandpal,
        Vadim Ksenofontov, and Claudia Felser}
\email{felser@uni-mainz.de}
\affiliation{
Institut f\"ur Anorganische und Analytische Chemie,
Johannes Gutenberg - Universit\"at, D-55099 Mainz, Germany.}

\author{Hong-Ji Lin}
\affiliation{National Synchrotron Radiation Research Center - NSRRC,
Hsinchu, 30076, Taiwan}

\author{Jonder Morais}
\affiliation{Instituto de Fisica, Universidade Federal do Rio Grande
do Sul, Porto Alegre, 91501-970, Brazil}

\date{\today}

\begin{abstract}
In this work a simple concept was used for a systematic search for
new materials with high spin polarization. It is based on two
semi-empirical models. Firstly, the Slater-Pauling rule was used for
estimation of the magnetic moment. This model is well supported by
electronic structure calculations. The second model was found
particularly for Co$_2$ based Heusler compounds when comparing their
magnetic properties. It turned out that these compounds exhibit
seemingly a linear dependence of the Curie temperature as function
of the magnetic moment.

Stimulated by these models, Co$_2$FeSi was revisited. The compound
was investigated in detail concerning its geometrical and magnetic
structure by means of X-ray diffraction, X-ray absorption and
M\"o\ss bauer spectroscopies as well as high and low temperature
magnetometry. The measurements revealed that it is, currently, the
material with the highest magnetic moment ($6\mu_B$) and
Curie-temperature (1100K) in the classes of Heusler compounds as
well as half-metallic ferromagnets. The experimental findings are
supported by detailed electronic structure calculations.
\end{abstract}



\maketitle

\section{Introduction}

Materials that exhibit half-metallic ferromagnetism are seen to be
potential candidates for the field of application being called
spintronics \cite{CVB02, ZFS04}, that is electronics making use of
the electron spin instead of its charge. The concept of
half-metallic ferromagnetism was first presented by de~Groot
\cite{GME83}, predicting it to appear in half Heusler compounds. The
model suggests that the density of states exhibits, around the Fermi
energy ($\epsilon_F$), a gap for minority electrons. Thus, these
materials are supposed to be 100\% spin polarized at $\epsilon_F$.
Most of the predicted half-metallic ferromagnets (HMF) belong to the
Heusler \cite{Heu03} compounds. In general, these are ternary
$X_2YZ$-compounds crystallizing in the $L2_1$ structure. $X$ and $Y$
are usually transition metals and $Z$ is a main group element.

Beside Heusler compounds, there are only few other materials being
predicted to be HMFs, like some oxides. Most of the predicted
compounds with Zincblende structure are chemically
uncharacterizable, at least up to now. Therefore, our research
concentrates on Heusler compounds.

High Curie temperatures, magnetic moments, and large minority gaps
are desirable for applications. For room temperature devices, in
particular, one needs to prevent a reduction of the spin
polarization and other magnetic properties by thermal effects. In
this context it should be noted that the Co$_2$ based Heusler
compounds exhibit the highest Curie temperature (985K, Co$_2$MnSi
\cite{BEJ83}) and the highest magnetic moment ($5.54\mu_B$ per unit
cell, Co$_2$FeGe \cite{BEJ83}) being reported up to now. The HMF
character of Co$_2$MnZ compounds was first reported by Ishida {\it
et al}\cite{IFK95}. Recently, Co$_2$MnSi \cite{KTH04} and Co$_2$MnGe
\cite{DAX05} were used to built first thin film devices. The present
work reports about Co$_2$FeSi. It will be shown that this compound
may be one of the most promising candidates for spintronic
applications.

\section{Theoretical Models}
\label{sec:TM}

There exist several rules counting the number of valence electrons
that are used to predict the magnetic moments in Heusler and other
magnetic compounds from the number of valence electrons. These rules
are based on particular numbers of valence elctrons starting with
the semi-conducting compounds where one has 18 valence electrons in
half Heusler compounds \cite{PST97}, or 24 in full Heusler
compounds. Rules that are not general but depend on the composition
of the sample are, somehow, not satisfactory. Therefore, it may be
helpful to start more basically. Pauling \cite{Pau38} pointed
already on the Heusler alloys in his basic work on the dependence of
the magnetic moments in alloys on the number of valence electrons
per atom.

\begin{figure}
\centering
\includegraphics[width=7.5cm]{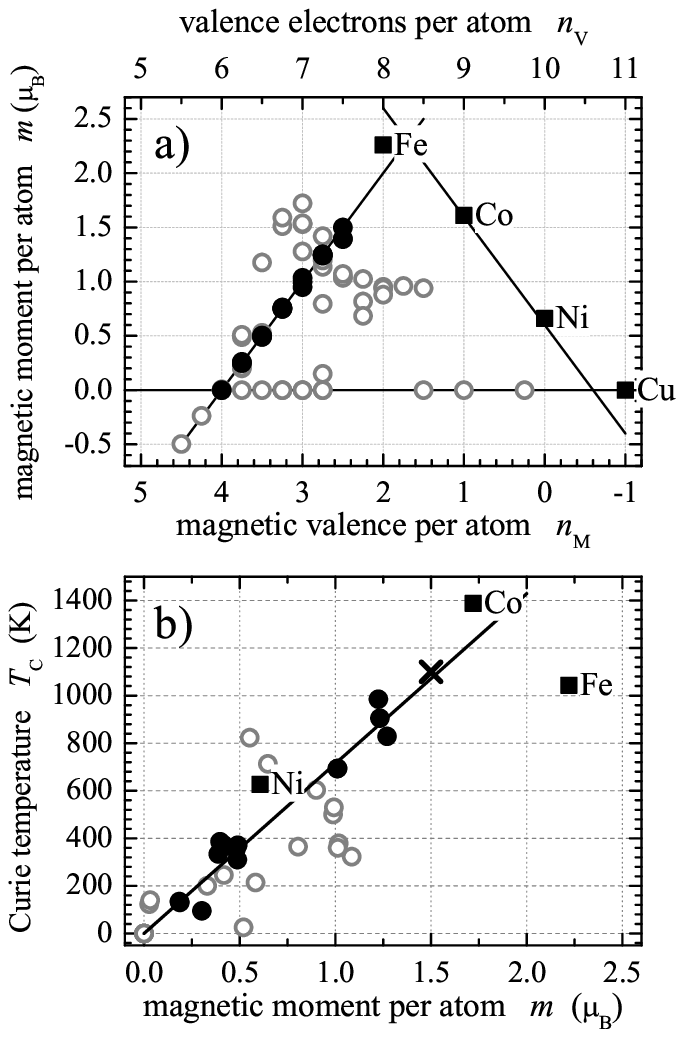}
\caption{Magnetic moments (a) and Curie temperature (b) of Heusler compounds.\\
         The heavy 3d transition metals are given for comparison. Full dots
         assign Co$_2$ based and open circles assign other Heusler
         compounds. The lines in a) assign the Slater-Pauling curve. The
         line in b) is found from a linear fit for Co$_2$ based compounds.
         The cross in b) assigns Co$_2$FeSi as measured in this work.}
\label{fig_1}
\end{figure}

Slater \cite{Sla36} and Pauling \cite{Pau38} found that the magnetic
moments ($m$) of $3d$ elements and their binary alloys can be
described by the mean number of valence electrons per atom ($n_V$).
The rule distinguishes the dependence of $m(n_V)$ into two regions
(see: Fig.\ref{fig_1}). In the second part ($n_V\geq8$), one has
closed packed structures (fcc, hcp) and $m$ decreases with
increasing $n_V$. This is called the region of itinerant magnetism.
According to Hund's rule, it is often favorable that the majority
$d$ states are fully occupied ($n_{d\uparrow}=5$). The magnetic
valence is defined by $n_M=2n_{d\uparrow}-n_V$, such that the
magnetic moment per atom is given by $m=n_M+2n_{sp\uparrow}$. A plot
of $m$ versus magnetic valence ($m(n_M)$) is called the generalized
Slater-Pauling curve (see Refs.\cite{MWM84, Kue84}). In the first
region ($n_V\leq8$), $m$ increases with increasing $n_V$. This part
is attributed to localized moments, where Fe (bcc) is a borderline
case. Here, the Fermi energy is pinned in a deep valley of the
minority electron density. This constrains the number of minority
$d$-electrons $n_{d\downarrow}$ resulting in $m=n_V-2
n_{\downarrow}$. For Fe and its binary alloys (Fe-Mn, Fe-Cr, and
partially Fe-Co) one has an average of approximately three minority
electrons occupied with the result $m\approx n_V-6$. Many Heusler
compounds exhibit an increase of $m$ with increasing $n_V$, and thus
they may belong to the first region of the Slater-Pauling curve.

Half-metallic ferromagnets (HMF) are supposed to exhibit a real gap
in the minority density of states and the Fermi energy is pinned
inside of the gap. From this point of view, the Slater-Pauling rule
is strictly fulfilled with
\begin{equation}
       m_{HMF}=n_V-6
\label{eq1}
\end{equation}
for the magnetic moment per atom, as the number of occupied minority
states $n_{\downarrow}$ has to be integer. The distribution of the
minority electrons on different bands, however, has to be found from
electronic structure calculations.

For ordered compounds with different kind of atoms it is, indeed,
more convenient to work with all atoms of the unit cell. In the case
of 4 atoms per unit cell, as in full Heusler (FH) compounds, one has
to subtract 24 (6 multiplied by the number of atoms) from the
accumulated number of valence electrons in the unit cell $N_V$ ($s,
d$ electrons for the transition metals and $s, p$ electrons for the
main group element) to find the magnetic moment per unit cell:
\begin{equation}
       m_{FH}=N_V-24.
\label{eq2}
\end{equation}

This {\it valence electron rule} \footnote{This term is used in the
following to distinguish the rule derived from the overall number of
valence electrons in the unit cell from the more general Slater
Pauling.} is strictly fulfilled for HMF only, as was first noted in
\cite{Kue84} for half Heusler (HH) compounds with 3 atoms per unit
cell ($m_{HH}=N_V-18$). In both cases the magnetic moment per unit
cell becomes strictly integer (in multiples of Bohr magnetons) for
HMF, what may be seen as an advantage of the {\it valence electron
rule} compared to the original Slater-Pauling law (Eq.\ref{eq1})
even so it suggests the existence of different laws.

The Slater-Pauling rule relates the magnetic moment with the number
of valence electrons per atom \cite{Sla36, Pau38}, but does not, per
se, predict a half-metallic behavior. The gap in the minority states
of Heusler compounds has to be explained by details of the
electronic structure (for examples see \cite{KWS83,GDP02} and
references there).

Self consistent field calculations were performed in order to
investigate the Slater-Pauling like behavior of Heusler compounds in
more detail (for details see Sec.\ref{sec:CD}). The electronic
structure of most known Heusler compounds was calculated in order to
find their magnetic moments and magnetic type. The calculations were
performed for overall 107 reported Heusler compounds from which 59
are based on $X$ and $Y$ being $3d$ metals, 17 with only $X$ and 28
with only $Y$ being a $3d$ metal. The remaining ones contain $4d$
and $4f$ metals on the $X$ and $Y$ sites, respectively. In addition
calculations were performed for 50 reported Half-Heusler compounds.

It turned out that nearly all (if not simple metals) Co$_2$ based
compounds should exhibit half-metallic ferromagnetism. It is also
found that the calculated magnetic moments of the Co$_2$ based
compounds follow the Slater-Pauling curve as described above (see
Fig.\ref{fig_1}a). Some small deviations are possibly caused by
inaccuracies of the numerical integration.

Inspecting the other transition metal based compounds, one finds
that compounds with magnetic moments above the expected
Slater-Pauling value are X=Fe based. Those with lower values are
either X=Cu or X=Ni based, with the Ni based compounds exhibiting
higher moments compared to the Cu based compounds at the same number
of valence electrons. Moreover, some of the Cu or Ni based compounds
are not ferromagnetic independent of the number of valence
electrons. Besides Mn$_2$VAl, only compounds containing both, Fe and
Mn, were found to exhibit HMF character with magnetic moments
according to the Slater-Pauling rule. No attempts have been made to
perform calculations for compounds that turn out to be chemically
uncharacterizable.

Inspecting the magnetic data of the known Heusler compounds in more
detail (see data and references in \cite{LB19C,LB32C}), one finds a
very interesting aspect. Seemingly, a linear dependence is obtained
for Co$_2$ based Heusler compounds when plotting the Curie
temperature ($T_C$) of the known, $3d$ metal based Heusler compounds
as function of their magnetic moment (see Fig.\ref{fig_1}b).
According to this plot, $T_C$ is highest for those half-metallic
compounds that exhibit a large magnetic moment, or equivalent for
those with a high valence electron concentration as derived from the
Slater-Pauling rule. $T_C$ is estimated to be above 1000K for
compounds with $6\mu_ B$ by an extrapolation from the linear
dependence.

Co$_2$FeSi was revisited as a practical test for the findings above.
This compound was previously reported to have a magnetic moment of
only $5.18 \mu_B$ per unit cell and a Curie temperature of above
$980K$ \cite{NBR77, Bus88}. One expects, however, $m=6\mu_B$ and
$T_C$ to be above 1000K, from the estimates given above.

\section{Calculational Details}
\label{sec:CD}

As starting point, self consistent first principle calculations were
performed using the linearized muffin tin orbital (LMTO) method
\cite{JA00}, as this method is very fast. Using the experimental
lattice parameter ($a_{exp}$=5.64\AA, see Sec.\ref{sec:SP}), the
results predicted Co$_2$FeSi to be a regular ferromagnet with a
magnetic moment of $5.08\mu_B$ per formula unit. The latter value is
much too small compared to the experimental one of $6\mu_B$ (see
Sec.\ref{sec:MP}).

More detailed calculations were performed to check if the too low
value is the result of a particular method or the parameterization
of the energy functional. The Korringa-Kohn-Rostocker method as
provided by H.~Akai\cite{AD85} was chosen as this program allows
calculations in the muffin tin (MT) and the atomic sphere (ASA)
approximations. Additionally, it provides the coherent potential
approximation (CPA) to be used for disordered systems. This method
was used to estimate the influence of disorder on the magnetic
structure.

The calculations were started with the most common parameterizations
of the exchange-correlation functional as given by Moruzzi, Janak,
and Williams \cite{MJW78} (MJW), von~Barth and Hedin \cite{BH72}
(vBH), and Vosco, Wilk, and Nussair\cite{VWN80,VW80} (VWN). The
generalized gradient approximation (GGA) was used in the form given
by Perdew {\it et al}\cite{PCV92}. To include non-local effects, the
VWN parametrization with additions by Perdew {\it et
al}\cite{P86,PW86,MS91} was used (PYVWN). The so-called exact
exchange is supposed to give the correct values for the gap in
semi-conductors using local density approximation. Here it was used
in the form given by Engel and Vosko\cite{EV93} (EV).

\begin{table}
\caption{Magnetic moments of Co$_2$FeSi calculated for spherical potentials.\\
         The calculations were carried out by means of KKR in muffin tin (MT)
         and atomic sphere (ASA) approximations for $a=5.64$\AA. All values
         are given in $\mu_B$. Total moments are given per unit cell and site
         resolved values are per atom.}
\begin{ruledtabular}
\begin{tabular}{lccclccc}
  KKR   &           & MT       &          &   &           & ASA      &   \\
        & $m_{tot}$ & $m_{Co}$ & $m_{Fe}$ &   & $m_{tot}$ & $m_{Co}$ & $m_{Fe}$ \\
  \hline
  MJW   & 5.03      & 1.19     & 2.69     &   & 5.17      & 1.27     & 2.68  \\
  vBH   & 4.88      & 1.15     & 2.64     &   & 5.03      & 1.22     & 2.62  \\
  VWN   & 4.99      & 1.18     & 2.68     &   & 5.15      & 1.26     & 2.67  \\
  GGA   & 5.22      & 1.22     & 2.85     &   & 5.18      & 1.27     & 2.69  \\
  EV    & 5.19      & 1.24     & 2.77     &   & 5.15      & 1.26     & 2.68  \\
  PYVWN & 5.03      & 1.19     & 2.72     &   & 5.67      & 1.25     & 3.23  \\
\end{tabular}
\end{ruledtabular}
\label{tab:tab1}
\end{table}

The results of the calculations using different approximations for
the potential as well as the parameterization of the
exchange-correlation part are summarized in Tab.\ref{tab:tab1}. The
calculated total magnetic moments range from $\approx4.9\mu_B$ to
$\approx5.7\mu_B$, thus they are throughout too low compared to the
experiment. They include, however, the value of $5.27\mu_B$ found in
Ref.\cite{GDP02} by means of the full potential KKR method.

In the next step, the full potential linear augmented plane wave
(FLAPW) method as provided by Wien2k \cite{bsm01} was used to
exclude that the observed deviations are due to the spherical
approximation for the potential (MT or ASA) as used in the above
methods.

First, the exchange-correlation energy functional being
parameterized within the generalized gradient approximation was
used. The energy convergence criterion was set to $10^{-5}$.
Simultaneously, the charge convergence was monitored and the
calculation was restarted if it was larger than $10^{-3}$. For
$k$-space integration, a $20\times20\times20$ mesh was used
resulting in 256 $k$-points in the irreducible part of the Brillouin
zone. In cases of doubt about convergence or quality of the
integration, the number of irreducible $k$-points was doubled.

The calculated magnetic moments for most known Heusler compounds are
summarized in Fig.\ref{fig_1} and discussed in Sec.\ref{sec:TM}. It
turned out, however, that the magnetic moment of Co$_2$FeSi is still
too small compared to the experimental value.

Comparing the result for Co$_2$FeSi, the magnetic moments found by
the different calculational schemes are very similar (compare
Tabs.\ref{tab:tab1} and \ref{tab:tab2}), implying that Co and Fe
atoms are aligned parallely independent of the method used. The
small, induced moment at the Si atom (not given in
Tab.\ref{tab:tab1}) is aligned anti-parallely to that at the
transition metal sites. As for KKR, the use of the EV
parameterization of the energy functional did not improve the
magnetic moment, the result was only $5.72\mu_B$.

\begin{table}
\caption{Magnetic moments of Co$_2$FeSi calculated for full symmetry potentials.\\
         The calculations were carried out by means of FLAPW for $a=5.64$\AA.
         All values are given in $\mu_B$. Total moments are given per unit
         cell and site resolved values are per atom. $m_{int}$ is the part of the
         magnetic moment that cannot be attributed to a particular site.}
\begin{ruledtabular}
\begin{tabular}{lccccc}
        & $m_{tot}$ & $m_{Co}$ & $m_{Fe}$  & $m_{Si}$  & $m_{int}$  \\
  \hline
  LDA (VWN)  & 5.59      & 1.40     & 2.87   & -0.01  & -0.07  \\
  LDA (EV)   & 5.72      & 1.45     & 2.94   & -0.003 & -0.11  \\
  LDA$+U$    & 6         & 1.54     & 3.30   & -0.13  & -0.25  \\
\end{tabular}
\end{ruledtabular}
\label{tab:tab2}
\end{table}

\begin{figure*}
\includegraphics[width=15cm]{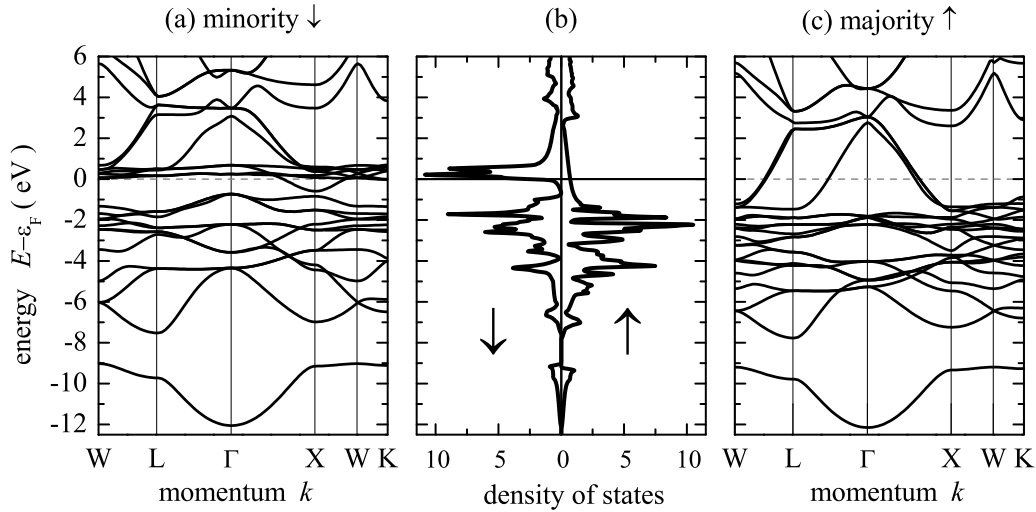}
\caption{LDA band structure and DOS of Co$_2$FeSi.\\
         The calculation was performed by Wien2k using the
         experimental lattice parameter.}
\label{fig_2}
\end{figure*}

Other than the LMTO or KKR methods (spherical potentials), the FLAPW
(full symmetry potential) calculations revealed a very small gap in
the minority states, but being located below the Fermi energy.
Figure \ref{fig_2} shows the band structure and density of states
calculated by FLAPW for the experimental lattice parameter.

The electronic structure shown in Fig.\ref{fig_2} reveals a very
small indirect gap in the minority states at about 2eV below
$\epsilon_F$. The fact that the Fermi energy cuts the minority bands
above the gap has finally the result that the magnetic moment is too
low and not integer, as would be expected for a half-metallic
ferromagnet.

\begin{figure}
\includegraphics[width=7.5cm]{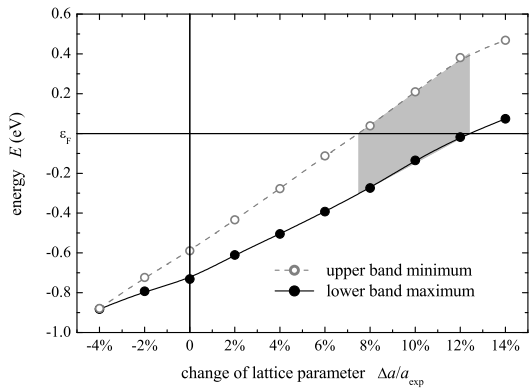}
\caption{Minority gap in Co$_2$FeSi.\\
         Shown are the positions of the extremal energies of the upper
         and lower bands depending on the lattice parameter ($a_{exp}=5.64$\AA).
         Energies are given with respect to the Fermi energy. The grey shaded area
         marks the domain of HMF character (lines are drawn to guide the eye).}
\label{fig_3}
\end{figure}

A structural refinement was performed to check if the experimental
lattice parameter minimizes the total energy. The dependence of the
energy with lattice parameter $a$ revealed that the minimum occurs
at the experimentally observed lattice parameter $a_{exp}$. From the
lattice parameter dependent calculations it showed up that the
experimentally found magnetic moment appears at larger values
of $a$. At the same time the size of the gap increased. Inspecting
the band structure, one finds that the Fermi energy is inside of the
gap for lattice parameters being enlarged by about 7.5\% to 12.5\%.
Figure \ref{fig_3} shows the dependence of the extremal energies of the
lower (valence) band and the upper (conduction) band of the minority
states enveloping the gap.

The magnetic moment is integer ($6\mu_B$) in the region were
$\epsilon_F$ falls into the gap (grey shaded area in
Fig.\ref{fig_3}), that is the region of half-metallic
ferromagnetism. The reason for the integer value is clear: the
number of filled minority states is integer and thus the magnetic
moment, too.

Usually, Heusler compounds are attributed to exhibit localized
moments. In that case, electron-electron correlation may play an
important rule. The LDA$+U$ scheme \cite{AAL97} was used for
calculation of the electronic structure to find out whether the
inclusion of correlation resolves the discrepancy between the
theoretical and measured magnetic moment. In Wien2k, the effective
Coulomb-exchange interaction ($U_{eff}=U-J$, where $U$ and $J$ are
the Coulomb and exchange parameter) is used to account for
double-counting corrections. It turned out that values of $U_{eff}$
from 2.5eV to 5.0eV for Co and simultaneously 2.4eV to 4.8eV for Fe
result in a magnetic moment of $6\mu_B$ and a gap in the minority
states.

\begin{figure*}
\includegraphics[width=15cm]{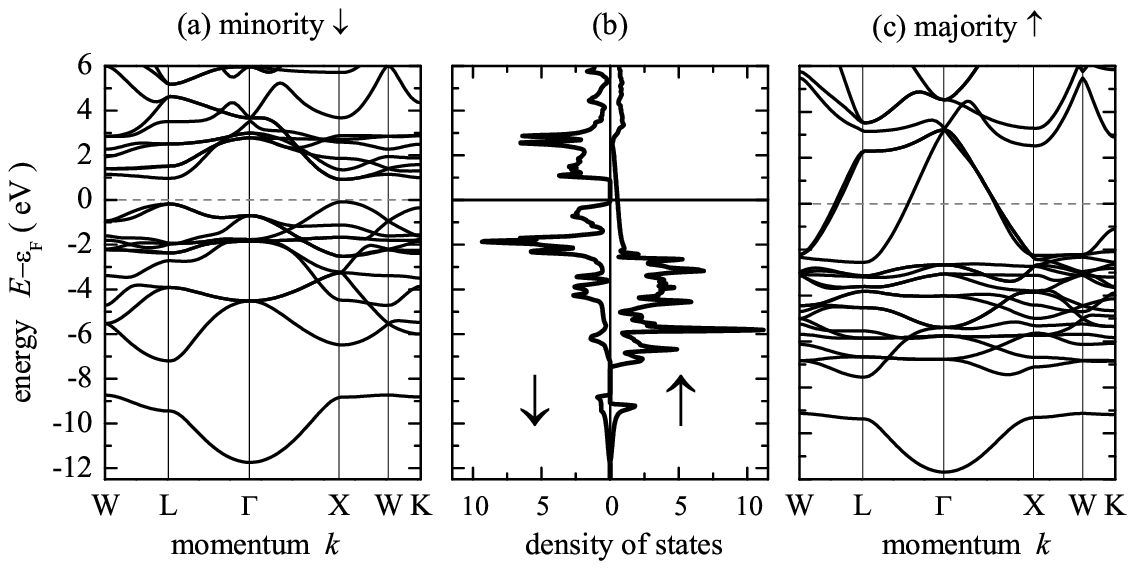}
\caption{LDA$+U$ band structure and DOS of Co$_2$FeSi.\\
         The calculation was performed by Wien2k using the experimental
         lattice parameter.}
\label{fig_4}
\end{figure*}

Figure \ref{fig_4} shows the band structure and density of states
calculated using the LDA$+U$ method. The effective Coulomb-exchange
parameter were set to U$_{eff,Co}$=4.8eV and U$_{eff,Fe}$=4.5eV at
the Co and Fe sites, respectively. These values are comparable to
those found in Ref.\cite{SS05} for bcc Fe (4.5eV) and fcc Ni (5eV).

The minority DOS (Fig.\ref{fig_4}) exhibits a clear gap around
$\epsilon_F$, confirming the half-metallic character of the
material. The high density below $\epsilon_F$ is dominated by
$d$-states being located at Co and Fe sites. Inspecting the majority
DOS one finds a small density of states near $\epsilon_F$. This
density is mainly derived from states located at Co and Si sites.

The distribution of the charge density on the atoms and states with
different orbital momentum is given in Tab.\ref{tab:tab3} (Note that
two Co atoms count for the sum of the charge in the unit cell.).

\begin{table}
\caption{Distribution of the valence states in Co$_2$FeSi.\\
         The number of occupied states was calculated by means of
         FLAPW using LDA$+U$ for $a=5.64$\AA. The muffin tin radii
         were set for all sites to $r_{MT}=1.22$\AA. This results in a space
         filling of 67\% by the atomic spheres, the remainder is taken
         as interstitial (int).}
\begin{ruledtabular}
\begin{tabular}{lclc}
                  & majority & & minority \\
  \hline
  Co               & 4.887   & & 3.343  \\
  Fe               & 5.063   & & 1.766  \\
  Si               & 1.247   & & 1.373  \\
  int              & 1.915   & & 2.176  \\
  \hline
  sum Co$_2$FeSi   & 18      & & 12  \\
\end{tabular}
\end{ruledtabular}
\label{tab:tab3}
\end{table}

From Tab.\ref{tab:tab3} it is found that in average three minority
states per atom are occupied ($2n_{\downarrow}=6$) as required by
the Slater-Pauling rule for the range of increasing magnetic moments
with an increasing number of valence electrons (see
Sec.\ref{sec:TM}). It is worthwhile to note that the same is true
for the other Heusler compounds shown in Fig.\ref{fig_1} exhibiting
half-metallic ferromagnetism. However, the electrons are distributed
in a different way across the X, Y, and Z sites.

\section{Experimental Details}

Co$_2$FeSi samples were prepared by arc-melting of stochiometric
quantities of the constituents in an argon atmosphere
($10^{-4}$mBar). Care was taken to avoid oxygen contamination. This
was established by evaporation of Ti inside of the vacuum chamber
before melting the compound as well as additional purification of
the process gas. Afterwards, the polycrystalline ingots were
annealed in an evacuated quartz tube at 1300K for 20 days. This
procedure resulted in samples exhibiting the correct Heusler type
$L2_1$ structure.

Flat disks were cut from the ingots and polished for spectroscopic
investigations at bulk samples. For powder investigations, the
remaining part was crushed by hand using a mortar. It should be
noted that pulverizing in a steel ball mill results in strong
perturbation of the crystalline structure.

X-ray photo emission (ESCA) was used to verify the composition and
to check the cleanliness of the samples. No impurities were detected
by means of ESCA after removal of the native oxide from the polished
surfaces by Ar$^+$ ion bombardment.

The geometrical structure was investigated by X-ray diffraction
(XRD) using excitation by Cu-K$_\alpha$ or
Mo-K$_\alpha$ radiation. Extended X-ray absorption
fine structure (EXAFS) measurements were performed at the {\it XAS}
beamline of LNLS (Campinas, Brazil) for additional structural
investigation, in particular to explain the site specific short
range order. Powder samples were investigated in transmission mode
using two ion chambers, bulk samples were investigated by the total
yield technique.

Magneto-structural investigations were carried out by means of
M\"o\ss bauer spectroscopy in transmission geometry using a constant
acceleration spectrometer. A $^{57}$Co(Rh) source with a line width
of 0.105mm/s was used for excitation.

The magnetic properties were investigated at low temperatures using
a super conducting quantum interference device (SQUID, Quantum
Design MPMS-XL-5). The high temperature magnetic properties were
investigated by means of a vibrating sample magnetometer (Lake Shore
Cryotronics, Inc., VSM Model 7300) equipped with a high temperature
stage. For site specific magnetometry, X-ray Magnetic
Circular Dichroism (XMCD) in photo absorption (XAS) was performed at
the {\it First Dragon} beamline of NSRRC (Hsinchu, Taiwan).

\section{Results}
\subsection{Structural Properties}
\label{sec:SP}

The correct $L2_1$ structure of the Co$_2$FeSi compound was verified
by XRD. The lattice constant was determined to be 5.64\AA. The
lattice parameter is obviously smaller than the one reported before
(compare Ref.\cite{NBH79}) and a lower degree of disorder is
observed in the present work.

A disorder between Co and Fe atoms ($DO_3$ type disorder) can be
excluded from the Rietveld refinement of the XRD data, as well as
from neutron scattering data (not shown here). A small ($<10$\%)
disorder between Fe and Si  atoms ($B2$ type disorder) can not be
excluded by either of these methods, particularly due to the low
intensities of the (111) and (200) diffraction peaks in XRD.

For further site specific structural information, EXAFS measurements
were carried out. A powder sample, as used for XRD, was investigated
in transmission mode.

\begin{figure}
\centering
\includegraphics[width=7.5cm]{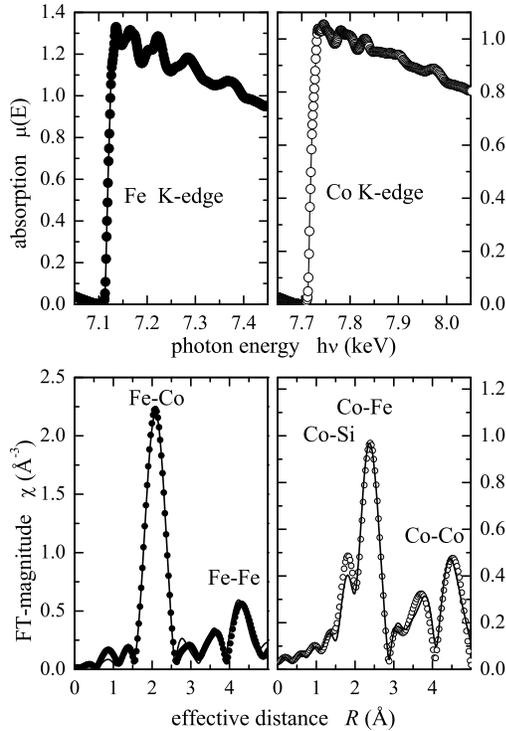}
\caption{EXAFS results for Co$_2$FeSi. \\
         The X-ray absorption spectra (with constant background removed)
         were taken at the K-edges of Co and Fe.
         The radial distribution functions derived from the spectra (symbols) are
         compared to the ones calculated (lines) using the best fit data.}
\label{fig_5}
\end{figure}

The absorption spectra collected at the Fe and Co K-edges are shown
in Fig.\ref{fig_5} after removal of a constant background. The ATOMS
\cite{RAV01} program was used to generate the structural input for
the EXAFS data analysis. The scattering parameter for all possible
scattering paths were calculated using FEFF6 \cite{ZRA95}. The
analysis of the EXAFS data was finally performed using the IFEFFIT
\cite{NRH95,New01} program package. The best fitting of the Fourier
transform modulus considering the $L2_1$ structure are also
displayed in Fig.\ref{fig_5}. It was not possible to fit the
experimental data to a structural model including $DO_3$ type
disorder, as expected. $B2$ type disorder can hardly be detected by
means of EXAFS as the distances in the first co-ordination shell of
Co and Fe are the same as in $L2_1$.

The radial distribution function $\chi(R)$ (FT magnitude) is produced
by forward Fourier transform of the EXAFS spectra after background
subtraction. It is given as function of the effective distance $R$.
Note that $R$ includes not only the interatomic distances $kR_j$ but
also the scattering phase shifts $\delta_j$ ($kR_j+\delta_j$). The
high intensity of $\chi(R)$ in the first co-ordination shell of Fe
points on the cubic environment consisting of eight Co atoms, as
expected for a well ordered Heusler compound. The peak of $\chi(R)$
for the first co-ordination shell of Co exhibits a clear splitting
being due to the different scattering phases of Fe and Si, although
the distance between these atoms and the Co is the same. The Fe
induced intensity of $\chi(R)$ is about half of that observed in the
Fe K-edge spectra. This confirms the tetragonal environment at the
Co sites with respect to Fe sites, as the scattering factors for
both types of atoms is nearly the same. Thus, the EXAFS measurements
corroborate the XRD results even at the short range order for both
Co and Fe.

\begin{figure}
\centering
\includegraphics[width=7.5cm]{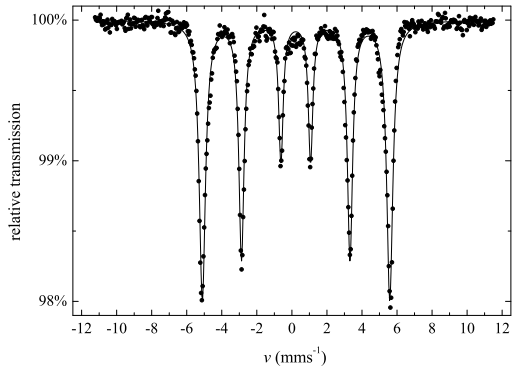}
\caption{M\"o\ss bauer-spectrum of Co$_2$FeSi. \\
         The spectrum was excited by $^{57}$Co and measured from a powder
         sample in transmission geometry at $T=85$K.}
\label{fig_6}
\end{figure}

For additional magneto-structural investigation, M\"o\ss bauer
spectroscopy was performed. The measurements were carried out at 85K
using powder samples. The observed 6-line pattern of the spectrum
(see Fig.\ref{fig_6}) is typical for a magnetically ordered system.
The observed $^{57}$Fe M\"o\ss bauer line width of 0.15mm/s is
characteristic for a well-ordered system. The value is comparable to
0.136mm/s observed from $\alpha$-Fe at 4.2K.

In detail, the M\"o\ss bauer spectrum shown in Fig.\ref{fig_6}
exhibits a sextet with an isomer shift of 0.23mm/s and a hyperfine
magnetic field of $26.3 \times 10^6$A/m. No paramagnetic line was
observed. Further, no quadrupole splitting was detected in
accordance with the cubic symmetry of the local Fe environment. A
$DO_3$ like disorder can be definitely excluded by comparing
measured and calculated hyperfine fields in ordered and disordered
structures.

\subsection{Magnetic Properties}
\label{sec:MP}

\begin{figure}
\centering
\includegraphics[width=7.5cm]{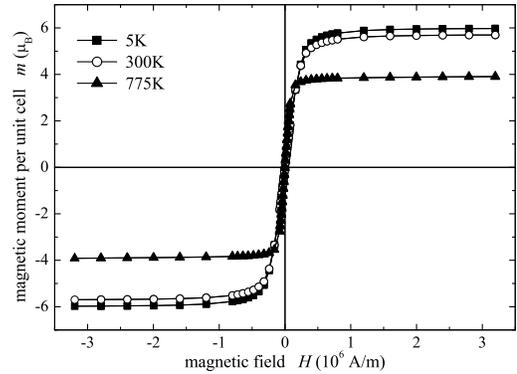}
\caption{Magnetic properties of Co$_2$FeSi. \\
         The field dependence of the magnetic moments was measured
         by SQUID magnetometry at different temperatures.}
\label{fig_7}
\end{figure}

Low temperature magnetometry was performed by means of SQUID to
proof the estimated saturation moment. The results are shown in
Fig.\ref{fig_7}. The measured magnetic moment in saturation is
$(5.97 \pm 0.05)\mu_B$ at 5K corresponding to $1.49\mu_B$ per atom.
An extrapolation to $6\mu_B$ per unit cell at 0K fits perfectly to
the one estimated from the Slater-Pauling rule. The measurement of
the magnetic moment reveals, as expected for a HMF, an integer
within the experimental uncertainty. Regarding the result of the
measurement (an integer) and the {\it valence electron rule}, it all
sums up to an evidence for half-metallic ferromagnetism in
Co$_2$FeSi. In more detail, Co$_2$FeSi turns out to be soft magnetic
with a small remanence of $\approx0.3$\% of the saturation moment
and a small coercive field of $\approx750$A/m, under the
experimental conditions used here. The magnetic moment for
Co$_2$FeSi was previously reported to be $5.90\mu_B$ \cite{NBR77} at
10.24K, but with a rather high degree of disorder (11\% $B2$ and
16\% $DO_3$). The same group reported later \cite{NBH79} a smaller
magnetic moment ($\approx5.6\mu_B$ interpolated to 0K) at a lattice
parameter of 5.657\AA, but still with a high degree of disorder.

The experimental magnetic moment is supported by the band structure
calculations, as was shown above, revealing a HMF character with a
magnetic moment of $6\mu_B$, if using appropriate parameters in the
self consistent field calculations.

\begin{figure}
\centering
\includegraphics[width=7.5cm]{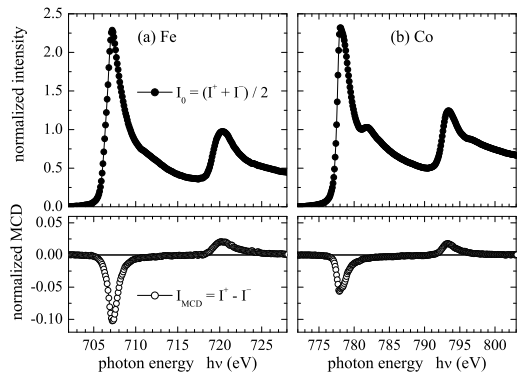}
\caption{Site resolved magnetic properties of Co$_2$FeSi. \\
         Shown are the XAS (I$_0$) and XMCD (I$_{MCD}$) spectra taken
         at the $L_{2,3}$ absorption edges of Fe (a) and Co (b) after
         subtracting a constant background.}
\label{fig_8}
\end{figure}

XMCD in photo absorption was measured to investigate the site
specific magnetic properties. The XAS and XMCD spectra taken at the
$L_{2,3}$ absorption edges of Fe and Co are shown in
Fig.\ref{fig_8}. The feature seen at 3eV below the $L_3$ absorption
edge of Co is related to the $L2_1$ structure and points on the high
structural order of the sample (it vanishes for $B2$ like disorder).
The magnetic moments per atom derived from a sum rule analysis
\cite{TCS92,CTA93} are $(2.6\pm0.1)\mu_B$ for Fe and
$(1.2\pm0.1)\mu_B$ for Co, at $T=300$K and $\mu_0H=0.4$T. The error
arises mainly from the unknown number of holes in the $3d$ shell and
the disregard of the magnetic dipole term in the sum rule analysis.
A pronounced enhancement of the orbital magnetic moments ($m_l$) as in Co$_2$Cr$_{1-x}$Fe$_x$Al \cite{EFV03}
or a field dependence of $m_l$ \cite{EWF04a} was not observed for Co$_2$FeSi.

The orbital moments were calculated using the LDA$+U$ scheme with
spin-orbit interaction (SO) and by LSDA calculations including the
Brooks orbital polarization term (OP) \cite{EJB89}. The LDA$+U$
calculations with SO revealed $r=m_l/m_s$ values of 0.05 and 0.02
for Co and Fe, respectively. These values are about a factor of two
higher than those found from LSDA + SO calculations without $U$. The
total magnetic moment stayed 6$\mu_B$ in this calculation as in
LDA$+U$ without SO. The OP calculations revealed slightly higher $r$
values. The orbital to spin magnetic moment ratios determined from
the XMCD measurements are about 0.05 for Fe and 0.1 for Co. All
values (ratio of Co and Fe moments, as well as values extrapolated
to 0K) are in good agreement to the electronic structure
calculations, keeping in mind that calculated site resolved values
depend always on the RMT-settings for the integration of the charge
density around a particular atom.

\begin{figure}
\centering
\includegraphics[width=7.5cm]{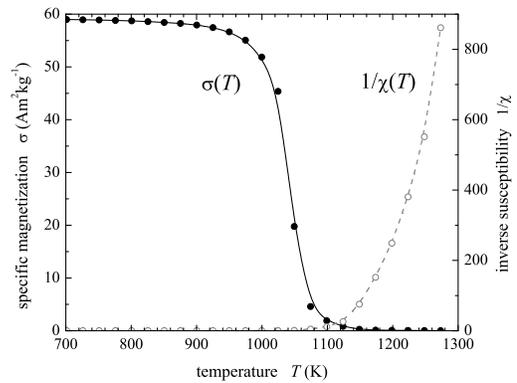}
\caption{Measurement of the specific magnetization as a function of
         temperature for Co$_2$FeSi. (Lines are drawn to guide the eye.}
\label{fig_9}
\end{figure}

The high temperature magnetization of Co$_2$FeSi was measured by
means of VSM. The specific magnetization as function of the
temperature is shown in Fig.\ref{fig_9}. The measurements were
performed in a constant induction field of $\mu_0H=0.1$T. For this
induction field, the specific magnetization at 300K is 37\% of the
value measured in saturation. The ferromagnetic Curie temperature is
found to be $T_C=(1100\pm20)$K. This value fits very well the linear
behavior shown in Fig.\ref{fig_1} for Co$_2$ based Heusler
compounds.

The Curie temperature is well below the melting point being obtained
by means of differential scanning calorimetry to be
$T_m=(1520\pm5)K$.
The paramagnetic Curie-Weiss temperature was identified to be
$\Theta\geq(1150\pm50)$K from the plot of the inverse susceptibility
$\chi^{-1}$ as a function of temperature $T$ by interpolating the
curve. The non-linearity in the $\chi^{-1}(T)$ dependence close to
$T_C$ indicates deviations from molecular field theory. A linear
dependence, however, is only expected at temperatures much higher
than $T_C$ that are not accessible here.

The highest known Curie temperature is reported for elemental Co to
be 1388K \cite{CRC01}. Only few materials exhibit a $T_C$ above
1000K, for example the Fe-Co binary alloys. With a value of
$\approx1100$K, Co$_2$FeSi has a higher Curie temperature than Fe
and the highest of all HMF and Heusler compounds being measured up
to now.

The present work leaves, however, some important questions open.
There is still no convincing theory that explains why the dependence
of $T_C$ on the magnetic moment should be linear over such a wide
range of compounds with different magnetic moments. For some alloys
a linear dependence is indeed found, but never over such a wide
range of different composition like for the Co$_2$ based HMF Heusler
compounds. An experimental challenge will be to find compounds with
magnetic moments of above $6\mu_B$ to prove whether it is possible
to find even higher Curie temperatures in this class of materials.
Co$_3$Si, which should exhibit $7\mu_B$, if following the {\it
valence electron rule}, crystallizes in a hexagonal but not in the
required Heusler structure, unfortunately.

\section{Summary and Conclusion}

In summary, it was shown how simple rules may be used to estimate
the properties of magnetic materials, in particular for those
Heusler compounds exhibiting half-metallic ferromagnetism.

As practical application of these simple rules, it was found that
the Heusler compound Co$_2$FeSi is a half-metallic ferromagnet
exhibiting the highest Curie temperature and magnetic moment. In
particular, a magnetic moment of $6\mu_B$ and a Curie temperature of
1100K were found in $L2_1$ ordered samples with a lattice parameter
of 5.64\AA.

The experimental findings are well supported by electronic structure
calculations. The comparison between experiment and calculations
gives clear advise that electron-electron correlation plays an
important role in Heusler compounds.

\begin{acknowledgments}
We thank Y.~Hwu (Academia Sinica, Taipei, Taiwan), M.~C.~M.~Alves
(UFRGS, Porto Alegre, Brazil), H.~Seyler, F.~Casper (Mainz), and the
staff of NSRRC (Hsinchu, Taiwan) and NLNS (Campinas, Brazil) for
help with the experiments. We acknowledge assistance of Lake Shore
Cryotronics, Inc. with the high temperature magnetic measurements.
Further, we thank G.~Frisch (Ludwig Alberts - University, Freiburg)
for performing Mo-K$_\alpha$ X-Ray diffraction.

This work is financially supported by DFG (research project FG 559),
DAAD (03/314973 and 03/23562), PROBRAL (167/04), and LNLS
(XAFS1-2372, XAFS1-3304).
\end{acknowledgments}


\end{document}